\documentclass[twocolumn]{aastex63}
\usepackage{graphics,epsf}
\usepackage{amsmath}                % American Mathematical Society package
\usepackage{amsfonts}               % American Mathematical Society fonts
\usepackage{amssymb}                % American Mathematical Society symbol
\usepackage{epsfig}                 % EPS figures
\usepackage{appendix}
\usepackage{graphicx}
\usepackage{float}
\usepackage{color}
\usepackage{multirow}
\usepackage{colortbl}
\usepackage[para,online,flushleft]{threeparttable}

\newcommand{\cm}{{~\rm cm}}

\newcommand{\g}{{~\rm g}}

\newcommand{\erg}{{~\rm erg}}
\newcommand{\yr}{{~\rm yr}}

\newcommand{\AU}{{~\rm AU}}

\newcommand{\days}{{~\rm days}}

\begin{document}

\title{Rapid expansion of red giant stars during core helium flash by waves propagation to the envelope and implications to exoplanets}

%\correspondingauthor{Ealeal Bear, Noam Soker}
%\email{}

\author{Ealeal Bear}
\affiliation{Department of Physics, Technion – Israel Institute of Technology, Haifa 3200003, Israel; ealealbh@gmail.com; soker@physics.technion.ac.il}

\author{Ariel Merlov}
\affiliation{Department of Physics, Technion – Israel Institute of Technology, Haifa 3200003, Israel; ealealbh@gmail.com; soker@physics.technion.ac.il}

\author{Yarden Arad}
\affiliation{Department of Physics, Technion – Israel Institute of Technology, Haifa 3200003, Israel; ealealbh@gmail.com; soker@physics.technion.ac.il}

\author[0000-0003-0375-8987]{Noam Soker}
\affiliation{Department of Physics, Technion – Israel Institute of Technology, Haifa 3200003, Israel; ealealbh@gmail.com; soker@physics.technion.ac.il}
\affiliation{Guangdong Technion Israel Institute of Technology, Guangdong Province, Shantou 515069, China}

\begin{abstract}
We assume that the strong convection during core helium flash of low mass red giant branch (RBG) stars excite waves that propagate to the envelope, and find that the energy that these waves deposit in the envelope causes envelope expansion and brightening. We base our assumption and the estimate of the waves’ energy on studies that explored such a process due to the vigorous core convection of massive stars just before they experience a core collapse supernova explosion. Using the stellar evolutionary code \textsc{mesa} we find that the waves’ energy causes an expansion within few years by tens to hundreds solar radii.   Despite the large brightening, we expect the increase in radius and luminosity to substantially enhance mass loss rate and dust formation. The dust shifts the star to become much redder (to the infrared), and the star might actually become fainter in the visible. The overall appearance is of a faint red transient event that lasts for months to few years. We suggest that in some cases envelope expansion might lead stars that are about to leave the RGB to engulf exoplanets. The extended envelope has a smaller binding energy to a degree that allows planets of several Jupiter masses or more and brown dwarfs to survive the common envelope evolution. We suggest this scenario to account for the planet orbiting the white dwarf (WD) WD~1856+534 (TIC~267574918) and for the WD - brown dwarf binary system ZTFJ003855.0+203025.5.
\end{abstract}


% ==========================================================
\section{Introduction} 
\label{sec:intro}
% ==========================================================

The presence of exoplanets in close orbits around evolved stars, i.e., horizontal branch stars or low mass white dwarfs (WDs) that are descendant of red giant branch (RGB) stars and WDs that are descendant of asymptotic giant branch (AGB) stars, is not easy to explain. By close orbits we refer to orbits with semi-major axes of $a \ll R_{\rm G}$ where $R_{\rm G}$ is the maximum radius that the giant progenitor has attained. Such close orbits imply that either the planet survived a common envelope evolution (CEE) or that it acquired its close orbit by dynamical interaction with a third body in the system. 

An example is the recently reported system  WD~1856+534 (TIC~267574918) where a planet orbits a WD with $a \simeq 0.02 \AU$ and an orbital period of $P_{\rm orb}=1.4 \days$ \citep{Vanderburgetal2020}. \cite{Vanderburgetal2020} claim that the large orbital separation makes the dynamical origin more likely that the CEE origin for this system. There are other claims for exoplanets orbiting WDs (e.g., \citealt{Gansickeetal2019, Manseretal2019}). 
\cite{JonesJenkins2014} refuted the claim of \cite{Setiawanetal2010} for an exoplanet orbiting a low metalicity horizontal branch star with an orbital period of $P_{\rm orb}=16.2 \days$ (for some other similar refuted claims see, e.g., \citealt{Krzesinskietal2020}). Nonetheless, the refuted claim of \cite{Setiawanetal2010} for an exoplanet orbiting a metal poor horizontal branch star that maintained a relatively massive envelope of $\simeq 0.3 M_\odot$ and at a relatively large orbital separation of $a \simeq 25 R_\odot$, promoted \cite{Bearetal2011} to speculate on a scenario where a metal-poor RGB star suffered a large expansion following its core helium flash and engulfed an exoplanet (section \ref{sec:Scenario}).

Systematic studies earlier than 2011 have already established the notion that planets can influence  the evolution on the RGB and beyond (e.g.,    \citealt{Soker1998, NelemansTauris1998, SiessLivio1999a, Carneyetal2003, DenissenkovHerwig2004, NordhausBlackman2006, Massarotti2008, SchroderSmith2008, Carlbergetal2009, VillaverLivio2009, Nordhausetal2010}; for later studies on RGB stars engulfing planets see, e.g., \citealt{Kunitomoetal2011, MustillVillaver2012, NordhausSpiegel2013, Villaveretal2014, AguileraGomezetal2016, Carlbergetal2016, Geieretal2016, Guoetal2016, Priviteraetal2016, Raoetal2018, Schaffenrothetal2019, Hegazietal2020, Jimenezetal2020, Krameretal2020}).   
However, without lowering the envelope binding energy just at the termination of the RGB, the planet that the RGB star engulfs spirals to very small orbits and either the RGB core tidally destroys the planet or the final orbit is very small and the star maintains only a very light envelope.
    
Regarding the planet-WD system WD~1856+534, one group of models for its formation considers the scattering of the planet to a close orbit around the WD by other planet(s) \citep{Maldonadoetal2021}, or by a another star i.e., the Lidov-Kozai effect (e.g., \citealt{MunozPetrovich2020, OConnoretal2021, Stephanetal2020, Vanderburgetal2020}). Another group of models considers the CEE. \cite{Lagosetal2021} study a CEE that takes place during the AGB phase of the WD progenitor. \cite{Chamandyetal2021} take the extra energy source that allows a planet to survive to come from the orbital energy that an inner planet releases as it enters the giant envelope first 
and causes its expansion (e.g., \citealt{SiessLivio1999a, SiessLivio1999b, Staffetal2016}). Inner planets can indeed help outer planets to survive the evolution of their parent star (e.g., \citealt{Bearetal2011, Lagosetal2021}). 
 
We describe the basic scenario and its assumptions in section \ref{sec:Scenario}. We then present our numerical scheme (section \ref{sec:MESA}) and the possible results of waves that the core helium flash excite in the core and propagate to the envelope  (sections \ref{sec:Case16Mo}). We summarize our results and discuss implications to exoplanets orbiting horizontal branch stars and WDs in section \ref{sec:summary}. 

% ==========================================================
\section{The proposed scenario}
\label{sec:Scenario}
% ==========================================================

\cite{Bearetal2011} considered the energy source that causes envelope expansion during the core helium flash to be the ignition of hydrogen at the base of the hydrogen-rich envelope in metal poor stars. They based their speculative scenario on the results of \cite{Mocaketal2010} who calculated hydrogen ignition by the core helium flash. In the calculations of \cite{Mocaketal2010} the hydrogen burning provides $\approx 1 \times 10^{48} \erg$ during the first year, i.e., an average luminosity of $L_{\rm H} \approx 10^7 L_\odot$ (their Figure 1). After a year this luminosity decreases to $L_{\rm H} \approx 10^6 L_\odot$, still much larger than the RGB luminosity.  
The huge energy production by the core convection and by the hydrogen burning decay in a time scale of $\approx 10-100 \yr$ (e.g., \citealt{Mocaketal2010}). \cite{Bearetal2011} justified the mixing of hot core material with the hydrogen-rich envelope by the metal-poor star they studied and/or by rapid core rotation due to an inner planet that spun-up the core.  

\cite{Bearetal2011} manually added an energy of $E_{\rm in} = 8.5 \times 10^{46} \erg$ just above the hydrogen-burning shell of their stellar model. This amounts to $7 \%$ of the energy that the hydrogen burning releases in the model of \cite{Mocaketal2010}. \cite{Bearetal2011} injected the energy in a time period of 7 years at an average power of $L_{\rm in} = 10^5 L_\odot$. 
\cite{Bearetal2011} found in their calculation that the outer radius of the convective zone in the envelope increases by a factor of about 4. After about $100 \yr$ the star shrinks back to its original radius. 

We raise here the possibility for another energy source. 
Based on the calculations of \cite{QuataertShiode2012} and \cite{ShiodeQuataert2014} for pre-supernova massive stars, we consider the possibility that the vigorous convection during the core helium flash excites waves that propagate into the envelope. \cite{QuataertShiode2012} and \cite{ShiodeQuataert2014} proposed a process by which a fraction of the energy in the gravity waves that the core convection excites in pre-supernova massive stars converts into sound waves as the gravity waves propagate into the envelope. The sound waves dissipate in the envelope. This energy deposition leads to envelope expansion (e.g., \citealt{McleySoker2014, Fuller2017}). \cite{McleySoker2014} study one case and find the pre-supernova model they use, of an initial mass of $15 M_\odot$, to expand by a factor of two, from about $1000 R_\odot$ to about $2000 R_\odot$.  
In that respect we note that \cite{Mocaketal2010} find that the convection during the core helium flash does excite gravity waves in the core. They also find that convection carries most of the energy that the nuclear reactions liberate. 

There are earlier studies that consider other roles that the internal gravity waves that the core helium flash excites play in stellar evolution. \cite{Schwab2020}, for example, consider the extra mixing that these waves induce near the core-envelope boundary during the helium flash. 
\cite{MillerBertolamietal2020}, as another recent example, 
study how the waves can cause periodic photometric variabilities in hot subdwarf stars. 
In this study we consider a different role that these propagating waves might play. 

We do not calculate the propagation of waves from the core to the envelope, as this requires a separate study. We simply take the same wave luminosity (power) as \cite{ShiodeQuataert2014} take. This wave energy is \citep{LecoanetQuataert2013}, 
\begin{eqnarray}
\begin{aligned}
L_{\rm wave,0} & \approx \mathcal{M}^{5/8} L_{\rm conv} 
 = 
2.7 \times 10^6 
\\ & \times \left( \frac{\mathcal{M}}{0.001} \right) ^{5/8}
\left( \frac{L_{\rm conv}}{2 \times 10^8 L_\odot} \right)  L_\odot ,
\label{eq:LwaveFrac}
\end{aligned}
\end{eqnarray}
where 
\begin{equation}
L_{\rm conv} (r) = 4 \pi r^2 v^3_{\rm conv}(r) \rho(r)
\label{eq:Lconv}
\end{equation}
is the luminosity that the convection carry at radius $r$ and $\mathcal{M} = v_{\rm conv}/c_{\rm s}$ is the Mach number of the convective motion of velocity $v_{\rm conv}$. In what follows we take the maximum value of $L_{\rm wave,0}$ at each time.  
For scaling we use typical values from \cite{Mocaketal2009} and \cite{Mocaketal2009}, in addition to our simulations that we describe later. 

In what follows we will consider much lower wave energies than what equation (\ref{eq:LwaveFrac}) gives, $L_{\rm w} \ll L_{\rm wave,0}$. 

% ==========================================================
\section{Numerical setting}
\label{sec:MESA}
% ==========================================================
We use \textsc{mesa-single} version 10398 \citep{Paxtonetal2011,Paxtonetal2013,Paxtonetal2015,Paxtonetal2018,Paxtonetal2019}. 
We follow the example of $1M~pre~ms~to~wd$ and only change the mass and the stopping condition.
The mass was changed to either $M=1.6M_\odot$ that we focus our study on, or to $M=1.2M_\odot$ or $M=2M_\odot$. We set the termination condition of this phase to the time when carbon nucleosynthesis starts ($star~C~mass~max~limit = 0.02$) which signifies that the He flash has already occurred.

After determining the maximum amount of energy and the duration of the wave that we assume convection excited (using the profile files produced by \textsc{mesa}, we apply equation \ref{eq:LwaveFrac}). We manually insert a fraction of this maximum energy in the src folder in the run-star-extra.f in the subroutine: subroutine energy-routine file, when we set the pointer of $other~energy$ to true.

We insert the wave luminosity $L_{\rm W}$ during four years in the outer zone of the envelope, either outer $0.2M_{\rm env}$, or $0.5M_{\rm env}$, or $0.8M_{\rm env}$. For each of these three cases we find the appropriate mass coordinate, m($0.2M_{\rm env}) =0.8619M_\odot$, $m(0.5M_{\rm env}) =0.6546 M_\odot$ and $m(0.8M_{\rm env}) =0.4473 M_\odot$ and insert the energy above that mass coordinate at a constant power per unit mass. We examine four cases of wave power as we describe in section \ref{sec:Case16Mo}.

% ==========================================================
\section{Core helium flash}
\label{sec:Case16Mo}
% ==========================================================

We follow the evolution of stellar models with initial masses of $M_{\rm ZAMS}=1.2 M_\odot$, $1.6 M_\odot$, and $2 M_\odot$, but focus on the $M_{\rm ZAMS}=1.6 M_\odot$ model. We follow each stellar model evolution in more details in the time period around its core helium flash. We define the time scale $t_w$ that we set to zero at the peak of the wave luminosity (see below). At the beginning of the core helium flash (just before the flash) of the $M_{\rm ZAMS}=1.6 M_\odot$ model (now an RGB star) its luminosity, radius, core mass, and envelope mass are, $L_{\rm b}=2027.74L_\odot$, $R_{\rm b}=128.4R_\odot$, $M_{\rm core,b}=0.45M_\odot$, and $M_{\rm env,b}=1.01 M_\odot$, respectively. We use the subscript `b' to indicate values just before the core helium flash. 

% ==========================================================
\subsection{The energy in the waves}
\label{subsec:WaveEnergy}
% ==========================================================

\begin{figure}[t]
\centering
\includegraphics[trim=3.6cm 8.5cm 1.0cm 8.0cm ,clip, scale=0.6]{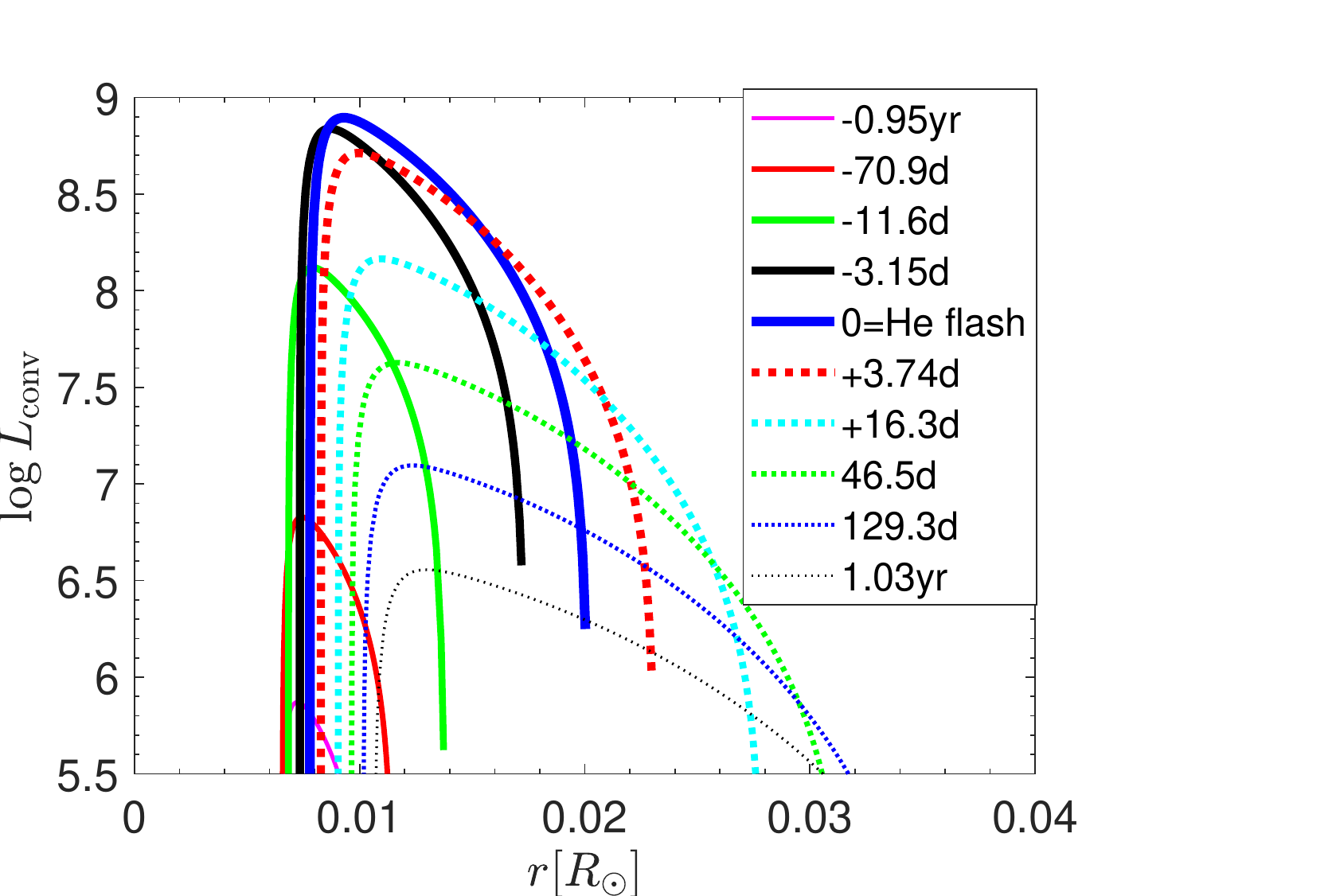} \\
 \includegraphics[trim=3.6cm 8.5cm 1.0cm 8.0cm ,clip, scale=0.6]{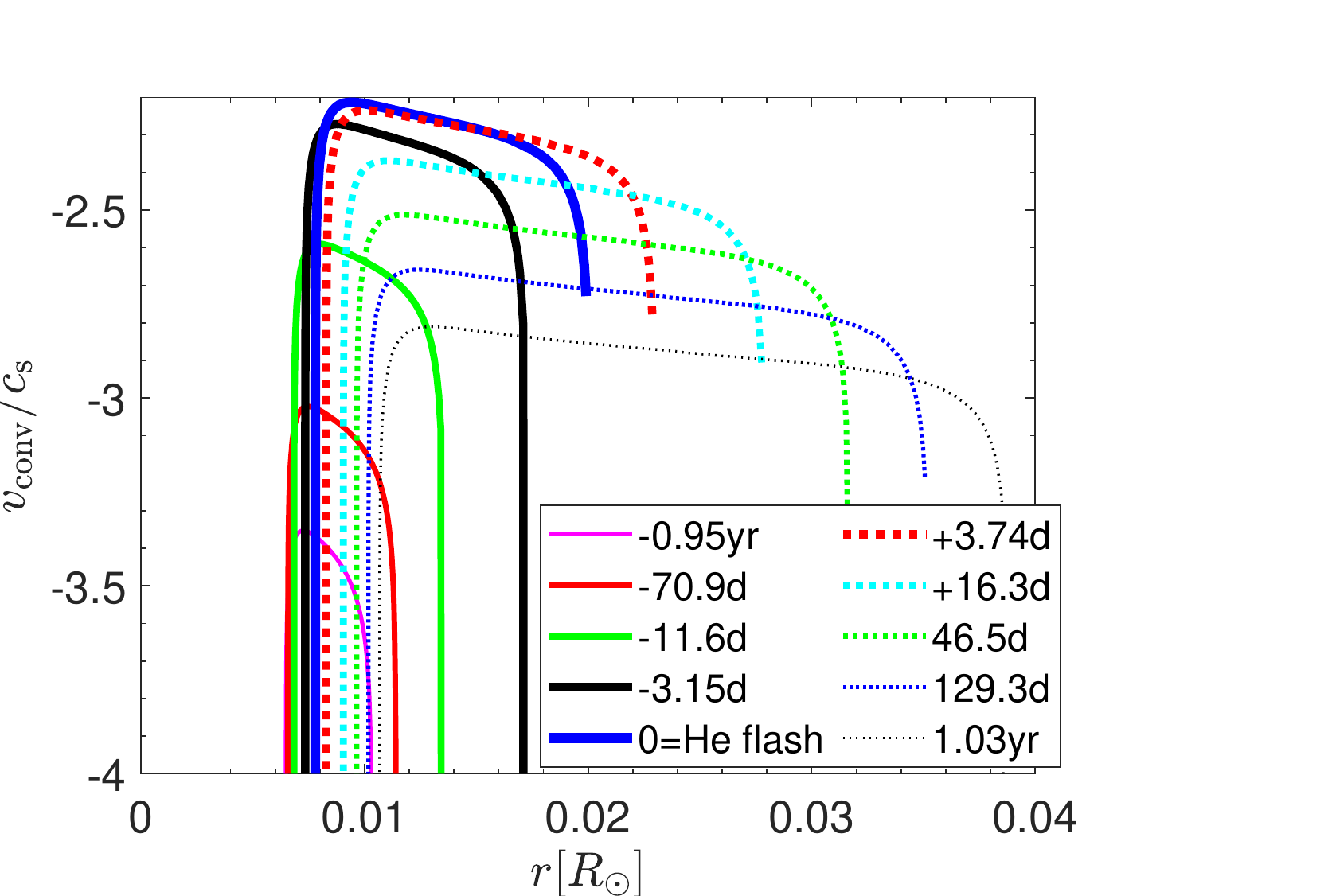} 
\caption{The convective luminosity (upper panel) and convective Mach number (middle panel) in the core of a stellar model with an initial mass of $M_{\rm ZAMS}=1.6 M_\odot$ versus radius, and at several times around the peak of the core helium flash. The times in the insets are $t_{\rm W}$ that is measured relative to the peak of the wave energy. Solid lines are for pre-peak and dotted line for post-peak values. The line get thinner away from the peak, both before and after the peak. }
 \label{fig:CoreHeliumFlash16}
 \end{figure}

The relevant properties of the core helium flash in our proposed scenario are the convective luminosity and the Mach number of the convective velocity (equation \ref{eq:LwaveFrac}).
In Fig. \ref{fig:CoreHeliumFlash16} we present the convective luminosity $L_{\rm conv} (r)$ according to equation (\ref{eq:Lconv}) (upper panel) the mach number of the convection ${\mathcal{M}} (r) = v_{\rm conv}/c_{\rm s}$ (lower panel) as function of radius in the core and at several times as indicated for a stellar model with an initial mass of $M_{\rm ZAMS}=1.6 M_\odot$.
% FFFFFFFFFFFFFFFFFFFFFFFFFFFFFFFFFFFFFFFFF
% FFFFFFFFFFFFFFFFFFFFFFFFFFFFFFFFFFFFFF

\begin{figure}[t]
% Instructions: [trim=Left,Bottom,Right,Up]
 \includegraphics[trim=3.6cm 8.5cm 1.0cm 8.0cm ,clip, scale=0.6]{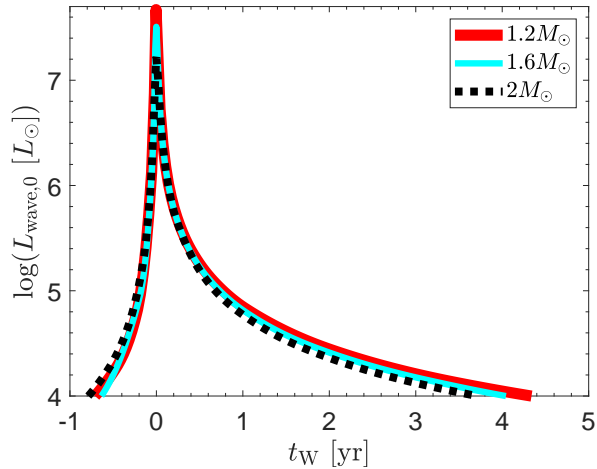} 
\caption{The wave luminosity according to equation (\ref{eq:LwaveFrac}) as function of time $t_{\rm W}$ measure from the peaks of eh wave energy for the three models of  $M_{\rm ZAMS}=1.2 M_\odot$, $M_{\rm ZAMS}=1.6 M_\odot$, and $M_{\rm ZAMS}=2 M_\odot$. We show only the time period when $L_{\rm wave,0} > 10^4 L_\odot$. The times of core helium flash for the three models are $t_{\rm He}=6.25 \times 10^9 \yr$, 
% $t_{\rm He}=6.2479408159340954 \times 10^9 \yr$, 
$t_{\rm He}=2.39 \times 10^9 \yr$,
%$t_{\rm He}=2.3927531863410258 \times 10^9 \yr$ 
and $t_{\rm He}=1.08 \times 10^9 \yr$ 
%$t_{\rm He}=1.0779580502585518 \times 10^9 \yr$ 
for  $M_{\rm ZAMS}=1.2 M_\odot$, $M_{\rm ZAMS}=1.6 M_\odot$, and $M_{\rm ZAMS}=2 M_\odot$, respectively. 
The masses of the cores at these times are $M_{\rm core,b}=0.415 M_\odot$, $M_{\rm core,b}=0.45 M_\odot$,  and $M_{\rm core,b}=0.43 M_\odot$, respectively.}
 \label{fig:LwaveFlash16}
 \end{figure}

In Fig. \ref{fig:CoreHeliumFlash16} we see the well known properties of the core helium flash that it has a distinct peak in time that last for several months and that the nuclear burning occurs off-center due to neutrino cooling in the center. In Fig. \ref{fig:LwaveFlash16} we plot the variation of $L_{\rm wave,0}(t)$ as function of time for the time period when $L_{\rm wave,0} > 10^4 L_\odot$. For comparison we also plot $L_{\rm wave,0}(t)$ for the two other stellar models that we simulate here.  
% FFFFFFFFFFFFFFFFFFFFFFFFFFFFFFFFFFFFFFFFF
% FFFFFFFFFFFFFFFFFFFFFFFFFFFFFFFFFFFFFF
 
Integrating over the wave power during the time period $\Delta t_4$ when $L_{\rm wave,0}>10^4 L_\odot$ we find the total wave energy for the $M_{\rm ZAMS}=1.6 M_\odot$ stellar model to be 
\begin{equation}
E_{\rm wave,0} = \int_{\Delta t_4} L_{\rm wave,0} dt = 2.1 \times 10^{47} \erg =1.7\times 10^6 L_\odot \yr. 
\label{eq:Ewave0}
\end{equation}
There is a large uncertainty in the variation of the power that the jets deposit to the envelope because the propagation time of the waves through the envelope is about the dynamical time, which is several months. For the present model $\Delta t_4 \simeq 4 \yr$. 
For that we consider the average luminosity over four years
\begin{equation}
L_{\rm W,0} \equiv \frac{E_{\rm wave,0}}{4 \yr} = 4.3 \times 10^5 L_\odot
\label{eq:LW0}
\end{equation}

For the cases of $M_{\rm ZAMS}=1.2M_\odot$ and $M_{\rm ZAMS}=2M_\odot$ the total energy in the waves are
about 20\% large and 20\% smaller than the value we give in equation (\ref{eq:Ewave0}), respectively. Since the envelope mass of the $M_{\rm ZAMS}=1.2M_\odot$ model is lower, it will suffer a much larger envelope expansion when waves dissipate their energy in the envelope with respect to the case we study here for $M_{\rm ZAMS}=1.6M_\odot$.

% ==========================================================
\subsection{Wave energy dissipation in the envelope }
\label{subsec:dissipation}
% ==========================================================

We do not know where exactly in the envelope the wave will deposit their energy. \cite{QuataertShiode2012} and \cite{ShiodeQuataert2014} proposed that the location of wave-energy deposition is in the outer regions where the maximum convective flux the envelope can carry, $L_{\rm max,conv}$ becomes below that of the wave power
\begin{equation}
L_{\rm max,conv} = 4 \pi r^2 c^3_{\rm s} \rho < L_{\rm wave}.
\label{eq:Lmax,conv}
\end{equation}
  
\begin{figure}[t]
% Instructions: [trim=Left,Bottom,Right,Up]
 \includegraphics[trim=3.6cm 8.5cm 1.0cm 8.0cm ,clip, scale=0.6]{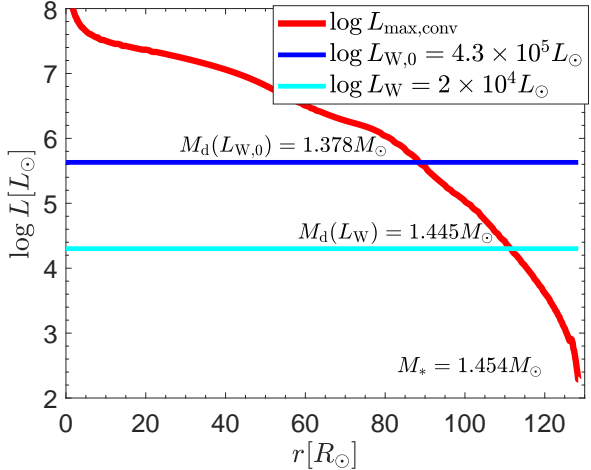} 
\caption{The maximum wave luminosity $L_{\rm max,conv}$ that convection can carry in the envelope according to equation (\ref{eq:Lmax,conv}) as function of radius for the $M_{\rm ZAMS}=1.6 M_\odot$ stellar model (red line). The blue line represents a wave energy of $L_{\rm W,0}$ according to equation (\ref{eq:LW0}) and the cyan line represents a wave energy of $L_{\rm W} = 2 \times 10^4 L_\odot$. Strong wave energy dissipation takes place in the outer zone where the wave energy is larger than $L_{\rm max,conv}$. We indicate the mass coordinates where wave energy equals $L_{\rm max,conv}$ and the total stellar mass.  }
 \label{fig:L_vs_R}
 \end{figure}

In Fig. \ref{fig:L_vs_R} we plot the variation of $L_{\rm max,conv}$ in the envelope and mark the values of $L_{\rm W,0}$ and of $L_{\rm W}=2 \times 10^4 L_\odot$ by horizontal lines. We plot the later value as we assume that the actual energy of the waves is much lower than $L_{\rm W,0}$. According to equation (\ref{eq:Lmax,conv}) and as we see in teh Fig. \ref{fig:L_vs_R} wave dissipation occurs in the outer parts of the envelope. By mass coordinate wave dissipation occurs in the zone $M>M_{\rm d}(L_{\rm W,0})=1.378 M_\odot$ for a wave power of $L_{\rm W,0}=4.3 \times 10^5 L_\odot$, and in the zone  $M>M_{\rm d}(L_{\rm W})=1.445 M_\odot$ for a wave power of $L_{\rm W,0}=2 \times 10^4 L_\odot$. The two zones correspond to envelope mass fractions of $7.5 \%$ and $0.9\%$, respectively. We will deposit the wave energy into much larger envelope zone, the outer $20\% - 80 \%$ of the envelope mass, to take into account the pressure that the wave exert as they propagate through the envelope before the radius of dissipation \citep{McleySoker2014}.   
% FFFFFFFFFFFFFFFFFFFFFFFFFFFFFFFFFFFFFFFFF
% FFFFFFFFFFFFFFFFFFFFFFFFFFFFFFFFFFFFFF

Because of the above uncertainty in the distribution of energy deposition, in section \ref{subsec:Outcomes} we present results for three simple prescriptions, where in each we spread the wave luminosity over the outer region of the envelope with a constant power per unit mass.  We spread the energy in either the outer $\xi=80\%$, $\xi=50\%$, or $\xi=20\%$ of the envelope mass. 

% ==========================================================
\subsection{Outcomes of wave-energy deposition}
\label{subsec:Outcomes}
% ==========================================================

In section \ref{subsec:WaveEnergy} we describe the possible properties of the waves that the convection excite during the core helium flash and their possible propagation and energy deposition to the envelope. Following the discussion there, we here present the results of energy deposition with a power of $L_{\rm W} \ll L_{\rm W,0}$ during a time period of $\Delta t_{\rm dep}=4 \yr$ for the model of $M_{\rm ZAMS}=1.6 M_\odot$.  
We deposit the energy to the envelope outer $\xi M_{\rm env,b}$ mass, with $\xi=80\%$, $\xi=50\%$, or $\xi=20\%$, and with a constant power per unit mass. We examine the response of the envelope for four different values of $L_{\rm W}$. Overall we have 12 energy deposition prescriptions. 

\begin{figure}%[ht]
 \includegraphics[trim=2.1cm 6.6cm 0.0cm 6.4cm ,clip, scale=0.52]{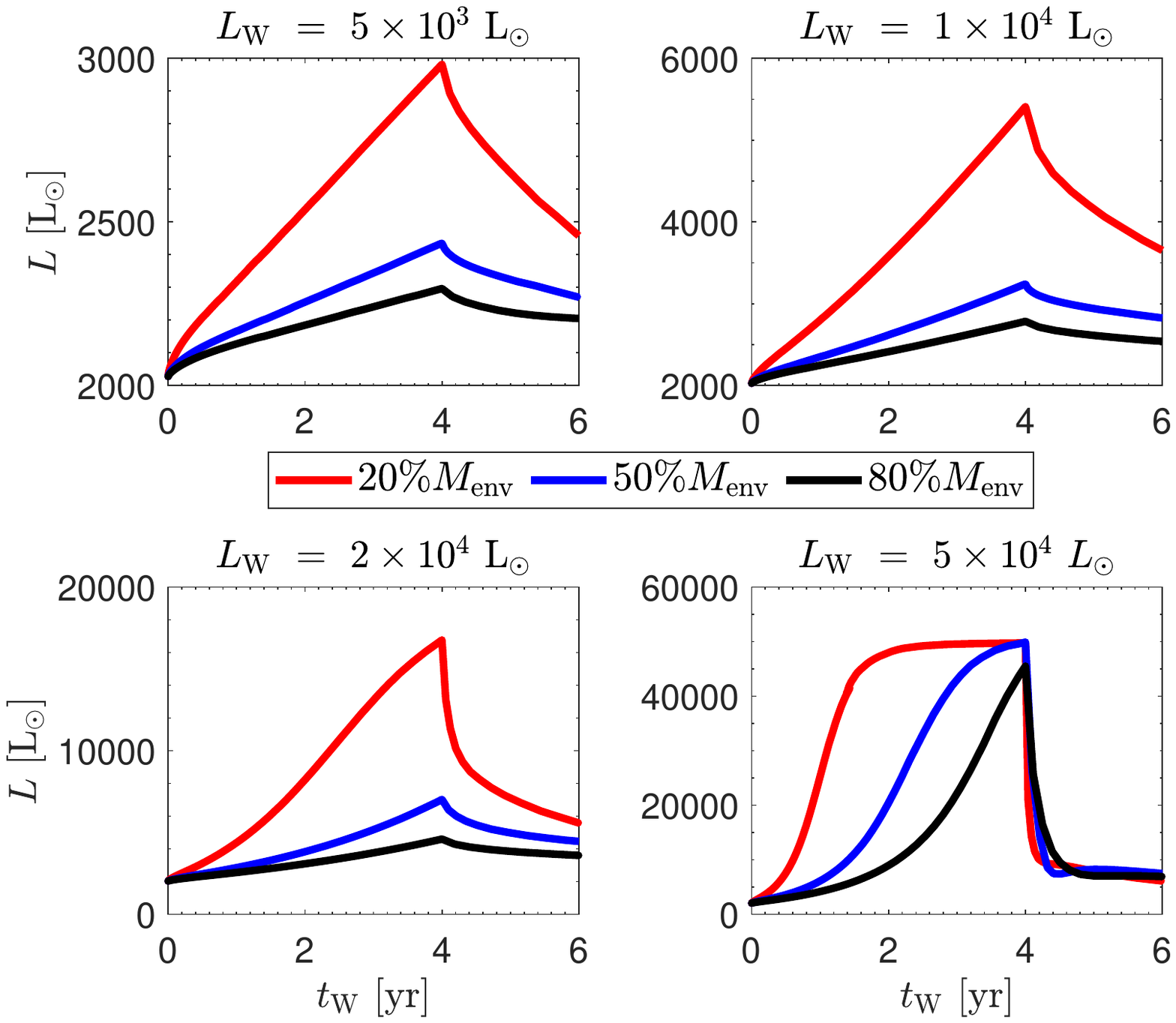} \\
 \includegraphics[trim=1.9cm 6.6cm 0.0cm 6.4cm ,clip, scale=0.52]{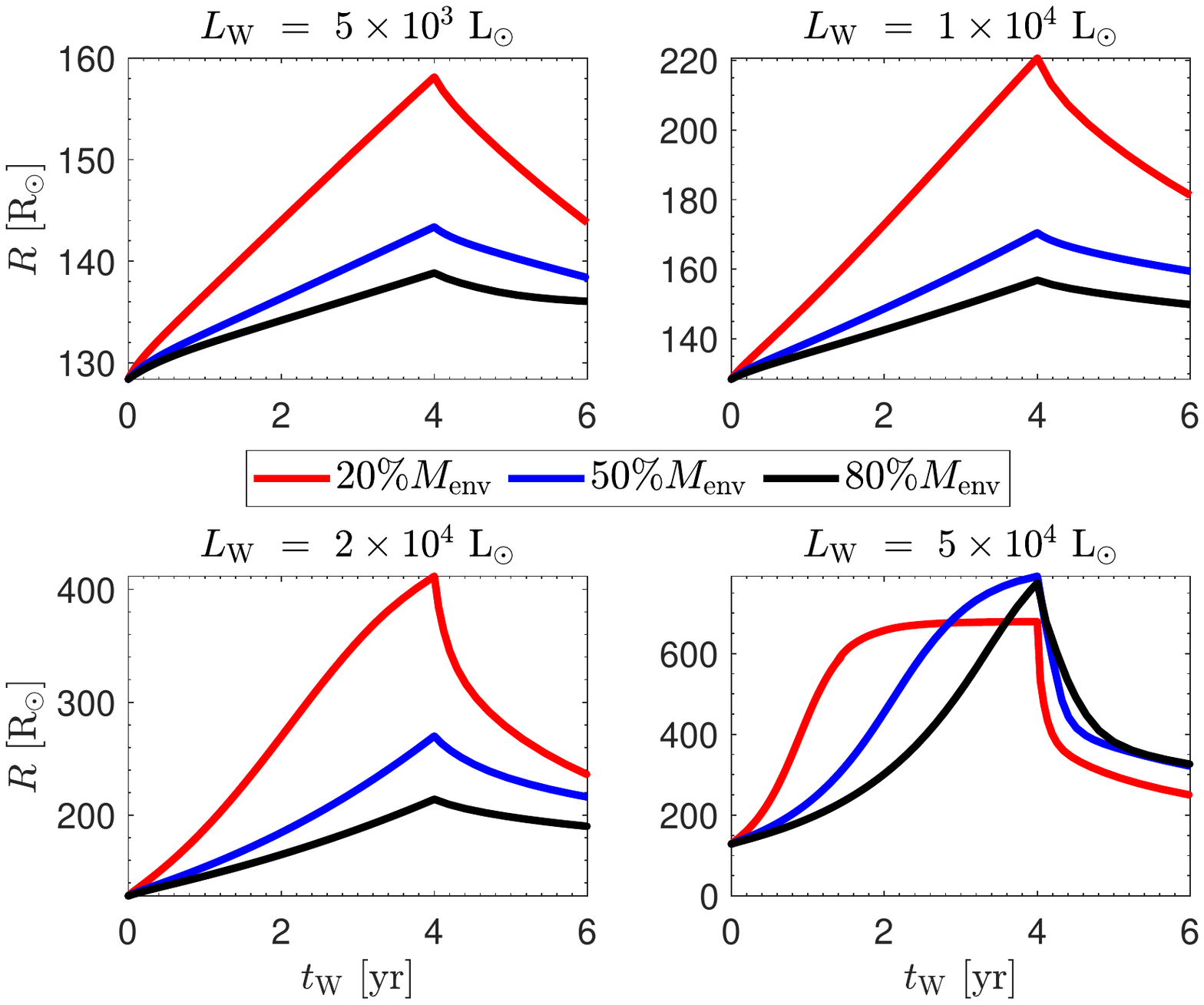}
\caption{The luminosity (upper two rows) and radius (lower two rows) as function of time during ($t_{\rm W}=0-4 \yr$) and after energy deposition as function of time $t_{\rm W}$ (set to zero at maximum wave luminosity). Each panel presents results for three different values of the outer envelope mass to which we deposit the energy. We indicate above each panel the power of the waves that we use. Note that the vertical scales and minimum values change from one panel to another. 
}
 \label{fig:RL_vs_time_M16.pdf}
 \end{figure}

\begin{figure}%[ht]
\includegraphics[trim=2.5cm 8.0cm 0.0cm 8.0cm ,clip, scale=0.55]{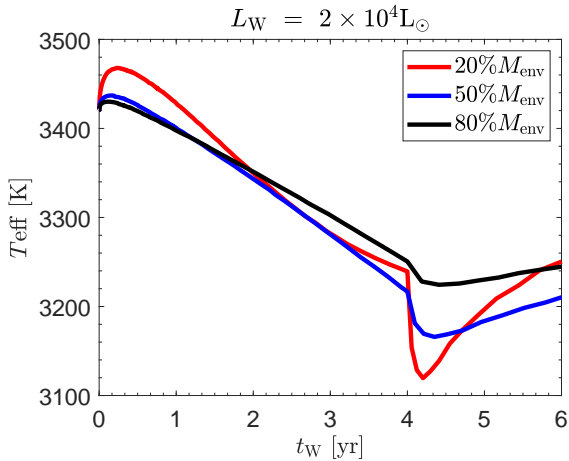}
\caption{Similar to the left panels in the second and fourth rows of Fig. \ref{fig:RL_vs_time_M16.pdf} with $L_{\rm W} =2 \times 10^4 L_\odot$, but for the effective temperature. }
 \label{fig:Teff_vs_time_E_2e4M16.pdf}
 \end{figure}

In Fig. \ref{fig:RL_vs_time_M16.pdf} we present the evolution of the luminosity and the stellar radius during and after the energy deposition for the 12 energy deposition prescriptions. In Fig. \ref{fig:Teff_vs_time_E_2e4M16.pdf} we present the evolution of the effective temperature for the cases with $L_{\rm W}=2 \times 10^4 L_\odot$. We see that the star relaxes towards its previous state on a time scale of several years. We will discuss the implications of this behavior below and in section \ref{sec:summary}.  
% FFFFFFFFFFFFFFFFFFFFFFFFFFFFFFFFFFFFFFFFF
  %[ht]
% FFFFFFFFFFFFFFFFFFFFFFFFFFFFFFFFFFFFFFFFF

% FFFFFFFFFFFFFFFFFFFFFFFFFFFFFFFFFFFFFFFFF
  %[ht]
% FFFFFFFFFFFFFFFFFFFFFFFFFFFFFFFFFFFFFF

\begin{figure}%[ht]
\includegraphics[trim=2.0cm 7.9cm 0.0cm 8.5cm ,clip, scale=0.52]{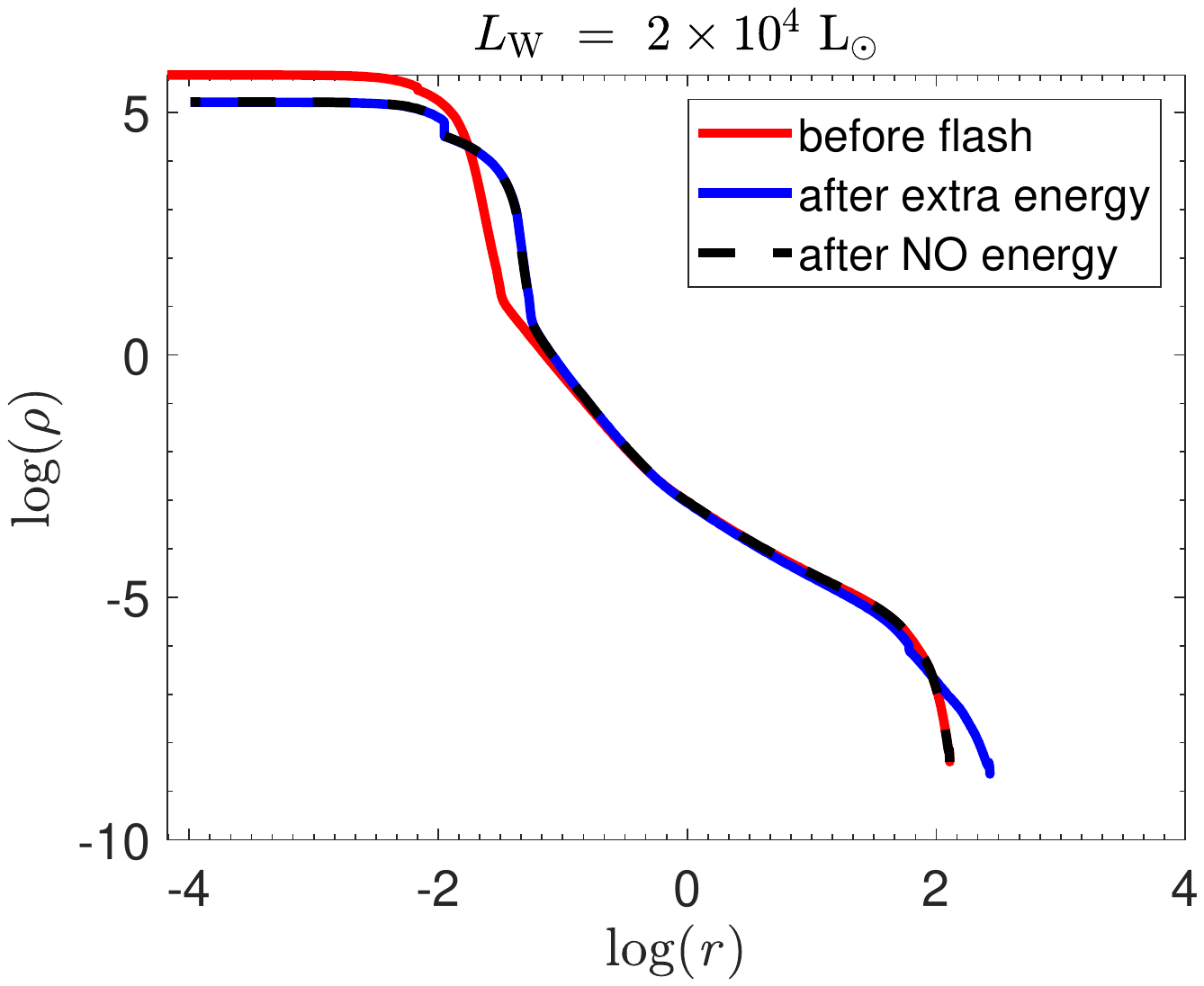} \\
\includegraphics[trim=2.0cm 7.9cm 0.0cm 8.5cm ,clip, scale=0.52]{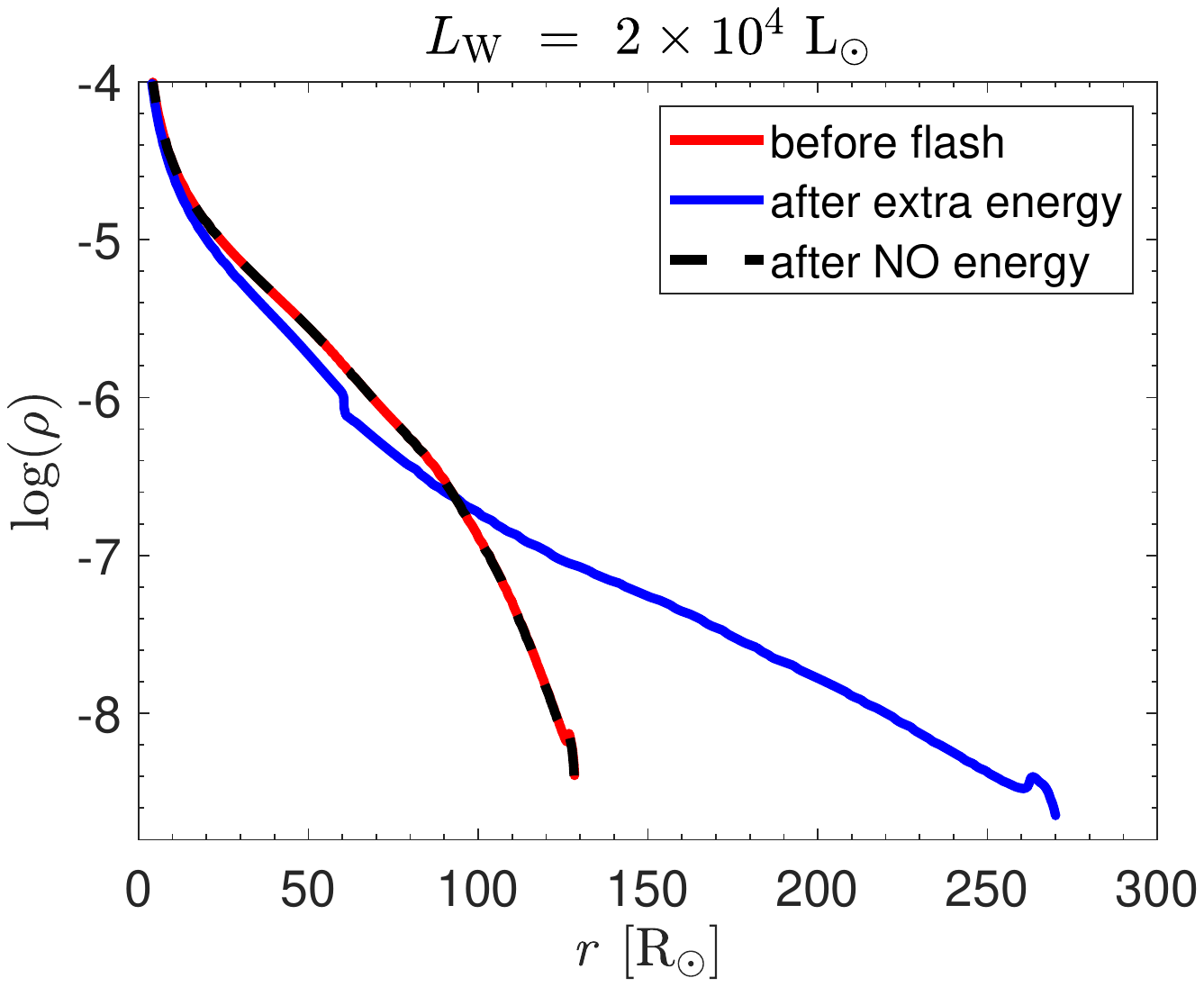}
\caption{The density (in units of $\g \cm^{-3}$) profiles of the case with $M_{\rm ZAMS}=1.6 M_\odot$, $L_{\rm W}= 2\times 10^4 L_\odot$ and  energy deposition in the outer $\xi = 50 \%$ envelope mass. We present the density profile before wave energy deposition (red line), at the end of wave energy deposition at $t_{\rm W}=4 \yr$, and at $t_{\rm W}=4 \yr$ but with no wave energy deposition.  The upper panel with logarithmic radius scale emphasizes the changes in the core due to the core helium flash, and the lower panel with a linear radius scale emphasizes the changes in the envelope due to wave energy deposition. 
}
 \label{fig:DensityProfiles16}
 \end{figure}

In Fig. \ref{fig:DensityProfiles16} we present the density profiles of the star just before the flash, at the end of the energy deposition time $t_{\rm W}=4 \yr$, and at $t_{\rm W}=4 \yr$ but without energy deposition. The upper panel emphasizes the changes in the core which are mainly due to the core helium flash and do not depend on wave energy deposition, while the lower panel emphasize the changes in the envelope that are due to wave energy deposition.  
% FFFFFFFFFFFFFFFFFFFFFFFFFFFFFFFFFFFFFFFFF
  %[ht]
% FFFFFFFFFFFFFFFFFFFFFFFFFFFFFFFFFFFFFFFFF

The wave energy dissipation in the envelope leads to a brightening of the star. The luminosity increase depends on the wave energy $L_{\rm W}$ and the mass of the envelope into which we deposit the wave energy. Here we deposit the energy into the outer zone, mass fraction $\xi$, of an envelope of mass $M_{\rm env,b} =1.01 M_\odot$ as we indicate in the different panels of Fig. \ref{fig:RL_vs_time_M16.pdf}.  The luminosity increases from about $L_{\rm b} = 2 \times 10^3 L_\odot$ to  $L(T_{\rm W}=4 \yr) = 2.2 \times 10^3 L_\odot$ in the case of $(L_{\rm W}, \xi)=(5 \times 10^3 L_\odot, 80\%)$ and to 
$L(T_{\rm W}=4 \yr) = 5 \times 10^4 L_\odot$ in the case of $(L_{\rm W}, \xi)=(5 \times 10^4 L_\odot, 20\%)$. Other values are in between. In the later case the envelope has reached a steady state and in the last three years of energy deposition ($t_{\rm W} \simeq 1 \yr$ to $t_{\rm W} = 4 \yr$) it emits all the wave energy. We see this also from the evolution of the radius of the star that reaches a constant value in the last three years of wave energy deposition. 
 
Consider the case of $(L_{\rm W}, \xi)=(2 \times 10^4 L_\odot, 20\%)$ as an example. Luminosity increases to $L=1.7 \times 10^4 L_\odot$ over four years. However, we do not expect the observation of such an increase in the visible band. After one year the luminosity increases from its initial value of $L_{\rm b}=2 \times 10^3 L_\odot$ to $L(1) \simeq 4 \times 10^3 L_\odot$ and the radius increases from $R_{\rm b} =128 R_\odot$ to about $R(1)=180 R_\odot$. This must lead to a substantial increase in mass loss rate and a copious amount of dust formation. 
The large amounts of dust and possible increase of the molecular opacity in the cooler upper atmosphere (as \citealt{Kravchenkoetal2021} found for Betelgeuse) can even lead to a decrease of the stellar brightness in the visible.
The RGB star turns to a luminous red transient. This event might be classified as a weak and red `gap event', i.e., a transient event with a peak luminosity between those of classical novae and typical supernovae. In this case it would be on the lower boundary of gap transients.  
The onset of a common envelope evolution of a low mass companion ($M_2 \simeq 0.3-0.5 M_\odot$) that enters the envelope of and RGB star or an AGB star might lead to a similar transient event that have an increase of luminosity by $\approx 10^4 L_\odot$ at peak and an event that lasts for several years, and accompanied by enhanced mass loss rate and dust formation. 
We expect present and future sky surveys to detect such weak-red gap transients, a small fraction of which might be RGB stars experiencing a core helium flash. 

% ==========================================================
\section{Discussion and Summary}
\label{sec:summary}
% ==========================================================

The basic assumption of our study is that the gravity waves that the strong convection excite during the core helium flash drive waves that propagate all the way to the envelope. We based this assumption on a similar process that  \cite{QuataertShiode2012} and \cite{ShiodeQuataert2014} proposed and studied for massive stars just before they experience a core collapse supernova explosion (section \ref{sec:Scenario}). Using their wave power relation to the core convection (equation \ref{eq:LwaveFrac}) we find the average wave power over $\Delta_4 \simeq 4 \yr$ to be $L_{\rm W,0} =4.3 \times 10^5 L_\odot$ (equation \ref{eq:LW0} for the $M_{\rm ZAMS}=1.6 M_\odot$ model, and Fig. \ref{fig:LwaveFlash16} for the three models). Under their assumption the waves deposit their energy in the outer envelope where the envelope convection cannot transport the wave power (Fig. \ref{fig:L_vs_R}). We did not study the wave propagation, but only their power and the effect of energy deposition into the envelope of the $M_{\rm ZAMS}=1.6 M_\odot$ stellar model over four years during its core helium flash.

We implemented a more conservative approach and deposited much lower energies than $E_{\rm wave,0}$ according to equation (\ref{eq:Ewave0}) and in a more extended envelope zone than what Fig. \ref{fig:L_vs_R} suggests. We present the response of the $M_{\rm ZAMS}=1.6 M_\odot$ envelope to energy deposition in Fig. \ref{fig:RL_vs_time_M16.pdf} (luminosity and radius), in Fig. \ref{fig:Teff_vs_time_E_2e4M16.pdf} (effective temperature) and in Fig. \ref{fig:DensityProfiles16} (density profile). We found the degree of envelope expansion and luminosity increase, and that the star relaxes back on a time scale of several years. 

We discuss two consequences of wave energy deposition that lead to envelope expansion and luminosity increase. These are the appearance of a gap transient event (with peak luminosity between classical novae and typical supernovae) and the possible engulfment of orbiting planets.

In section \ref{subsec:Outcomes} we discussed our expectation that a small fraction of weak and red gap transients be RGB stars experiencing a core helium flash. Because of the large amount of dust that we expect the expanding star to form at the beginning of the event, the object might not increase its luminosity in the visible, and might actually become fainter in the visible. Such events are better observed in the infrared. 
 
We turn to discuss the possible engulfment of planets, that we further elaborate on in an accompanying paper \citep{Merlovetal2021}. As we show in \cite{Merlovetal2021}, the expanding envelope as a result of wave energy deposition at the core helium flash might engulf planets in the right orbital separation that without the envelope expansion would survive the entire RGB phase of their parent star. This formation of a common envelope evolution brings the planet to spiral-in towards the core of the RGB star. Because of the inflated envelope its binding energy is lower, and the planet might unbind most of the envelope and therefore might survive the common envelope evolution.   

Consider the case of $(L_{\rm W}, \xi)=(2 \times 10^4 L_\odot, 50\%)$ for the $M_{\rm ZAMS} = 1.6 M_\odot$ stellar model. 
The binding energy of the envelope residing above radius $r= 1 R_\odot$ at the end of energy deposition at $t_{\rm W}=4 \yr$ is $E_{\rm e,bind} (1 R_\odot)=7\times 10^{45} \erg$.
We can compare this energy to the orbital energy that a planet of mass $M_{\rm p}$ releases as it spirals-in to an orbital separation of $a=1R_\odot$, $E_{\rm orbit} = 8 \times 10^{45} (M_{\rm p}/10 M_{\rm J}) \erg$, where $M_{\rm J}$ is Jupiter mass. Because  $E_{\rm orbit} \ga E_{\rm e,bind}$ it is possible for a planet to expel most of the envelope and survive such an evolution. Only more accurate numerical simulations of common envelope evolution with planets (e.g., \citealt{Krameretal2020}) can determine whether the planet survives or not. 

In \cite{Merlovetal2021} we suggest that the general scenario that we discussed in this paper might explain the system WD~1856+534 (TIC~267574918) where a planet orbits a WD of mass $M_{\rm WD} \simeq 0.52 M_\odot$ with $a \simeq 0.02 \AU$ and an orbital period of $P_{\rm orb}=1.4 \days$ \citep{Vanderburgetal2020}. We discussed some other scenarios for the formation of this system in section \ref{sec:intro}, and present the evolution towards a common envelope in \cite{Merlovetal2021}. Here we note that in our models of  $(L_{\rm W}, \xi)=(2 \times 10^4 L_\odot, 50\%)$ and $(L_{\rm W}, \xi)=(2 \times 10^4 L_\odot, 20\%)$  for the $M_{\rm ZAMS} = 1.6 M_\odot$ stellar model the binding energies of the envelope that resides above mass coordinate $m=0.52 M_\odot$ at the end of wave energy deposition is $E_{\rm e,bind} (0.52M_\odot) =3.9\times 10^{45} \erg$ and $E_{\rm e,bind} (0.52M_\odot) =6.4\times 10^{45} \erg$, respectively. Depositing the wave energy in a lower envelope mass fraction (in our study 20\% of the envelope mass) leads to a larger envelope expansion that allows the envelope to engulf more exoplanets. However, because of larger emission the decrease in the envelope binding energy is less pronounced. The binding energy without wave energy deposition is much larger $E_{\rm e,bind,b} (0.52M_\odot) =1.2\times 10^{46} \erg$. 
The orbital energy that the planet releases as it spirals-in to an orbital separation of $a=4 R_\odot$ around that WD is $E_{\rm orbit} = 2.5 \times 10^{45} (M_{\rm p}/10 M_{\rm J}) \erg$. For the above scaling, the planet might directly  unbind a large fraction of the envelope mass. A lower mass progenitor, e.g., $M_{\rm ZAMS} \simeq 1.2-1.4 M_\odot$, makes the scenario more favorable. 

If the RGB star engulfs inner planets to the planet that eventually survives, the inner plant(s) can increase the likelihood of the planet to survive. \cite{SiessLivio1999a, SiessLivio1999b} show that accretion process of a planet onto the core is accompanied by a substantial expansion of the star that can lead to high mass ejection. Namely, an inner planet or two can substantially reduce the mass of the RGB envelope when the surviving planet enters the envelope during the core helium flash. 

Finally, we also suggest that the scenario we studied here where convection-excited waves cause large envelope expansion during core helium flash might explain the formation of the system ZTFJ003855.0+203025.5 that \cite{vanRoesteletal2021} discovered, where a brown dwarf of mass $\simeq 0.059M_\odot$ and a WD of mass $\simeq 0.5 M_\odot$ orbit each other with a semi-major axis of $2.0 R_\odot$.

%%%%%%%%%%%%%%%%%%%%%%%%%%%%%%%%%%%%%%%%%%%%%%%%
\acknowledgments
%%%%%%%%%%%%%%%%%%%%%%%%%%%%%%%%%%%%%%%%%%%%%%%%
This research was supported by a grant from the Israel Science Foundation (769/20)

%%%%%%%%%%%%%%%%%%%%%%%%%%%
\textbf{Data availability}

The data underlying this article will be shared on reasonable request to the corresponding author. 
%%%%%%%%%%%%%%%%%%%%%%%%%%%

%\software{MESA \citep{Paxtonetal2011, Paxtonetal2013, Paxtonetal2015,Paxtonetal2018, Paxtonetal2019}}

%\pagebreak

\end{document}